\documentclass[11pt,twoside]{article}
\usepackage{asp2010}

\resetcounters
\begin{document}

\title{On the nature of the stellar bridge between Leo IV and Leo V}
\author{Shoko Jin$^{12}$, Nicolas Martin$^{32}$, Jelte de Jong$^4$, Blair Conn$^{2}$, Hans-Walter Rix$^{2}$ and Mike Irwin$^5$
\affil{$^1$Astronomisches Rechen-Institut, Zentrum f\"ur Astronomie der Universit\"at Heidelberg, M\"onchhofstr. 12--14, D-69120 Heidelberg, Germany}
\affil{$^2$Max-Planck-Institut f\"ur Astronomie, K\"onigstuhl 17, D-69117 Heidelberg, Germany}
\affil{$^3$Observatoire Astronomique de Strasbourg, CNRS, UMR 7550, 11 rue de l'Universit\'e, F-67000 Strasbourg, France}
\affil{$^4$Leiden Observatory, Leiden University, P.O. Box 9513, 2300 RA Leiden, The Netherlands}
\affil{$^5$Institute of Astronomy, University of Cambridge, Madingley Road, Cambridge CB3 0HA, UK}}

\begin{abstract}

We present a revised analysis of a speculated stellar bridge between the Milky Way dwarf galaxies Leo IV and Leo V. Using data acquired with Subaru/Suprime-Cam over a $1\null^\circ\times4\null^\circ$ field encompassing the two satellites and the region in between, we confirm our previous detection of a stellar overdensity between Leo IV and Leo V \citep{2010ApJ...710.1664D}. The larger area coverage and improved depth of our current dataset allow for an improved analysis of the stellar overdensity that had previously appeared to bridge the two galaxies. A main-sequence turn-off feature visible in the stacked colour-magnitude diagram of the contiguously observed Subaru fields reveals an extended stellar structure at a distance of approximately 20 kpc. Its angular proximity to the Virgo overdensity, as well as a good correspondence in distance and metallicity, suggests that the smaller structure we detect may be associated with the much larger Virgo stellar overdensity.
\end{abstract}

\section{Introduction}
\label{sec:intro}
The discovery of numerous dwarf galaxies through the Sloan Digital Sky Survey (SDSS)\nocite{2000AJ....120.1579Y} over the last decade has helped to advance our understanding of the Milky Way in a cosmological context, with our appreciation of the Galaxy as a micro-cosmos, together with our ever-growing understanding of the Andromeda galaxy, leading to the development of `near-field cosmology'. The constellation that boasts the largest number of Milky Way satellites is the realm of Leo, and two of its five dwarf-galaxy inhabitants form the subject of our current study. Discovered by \cite{2007ApJ...654..897B,2008ApJ...686L..83B}, Leo IV and Leo V are close neighbours not just in projection --- separated on the sky by a mere $3\null^\circ$ --- but also spatially, sharing residence of the Galactic halo beyond 150~kpc. Targeted observations for studying these Milky Way satellites and their surrounding environment by \cite{2010ApJ...710.1664D} using the LAICA imager on the 3.5m telescope at the Calar Alto observatory led to the unexpected detection of an overdensity of stars, seemingly bridging the two galaxies. However, the loss of one of three planned fields due to bad weather unfortunately left us with only a tentative detection of the possible stellar bridge, whose true nature unfortunately could not be resolved from these data alone. Follow-up observations using Subaru/Suprime-Cam aimed at obtaining a wider and deeper dataset has enabled us to shed clearer light on the nature of this stellar overdensity, as well as allowing for an improved set of structural parameters to be derived for Leo IV and Leo V. These results are presented in summarised form in this contribution, with full details to be presented in a paper currently in preparation.

\section{Structural parameters of Leo IV and Leo V}
\label{sec:parameters}

\begin{figure}
\vspace{-0.2cm}
\begin{center}
\resizebox{0.9\textwidth}{!}{
\includegraphics{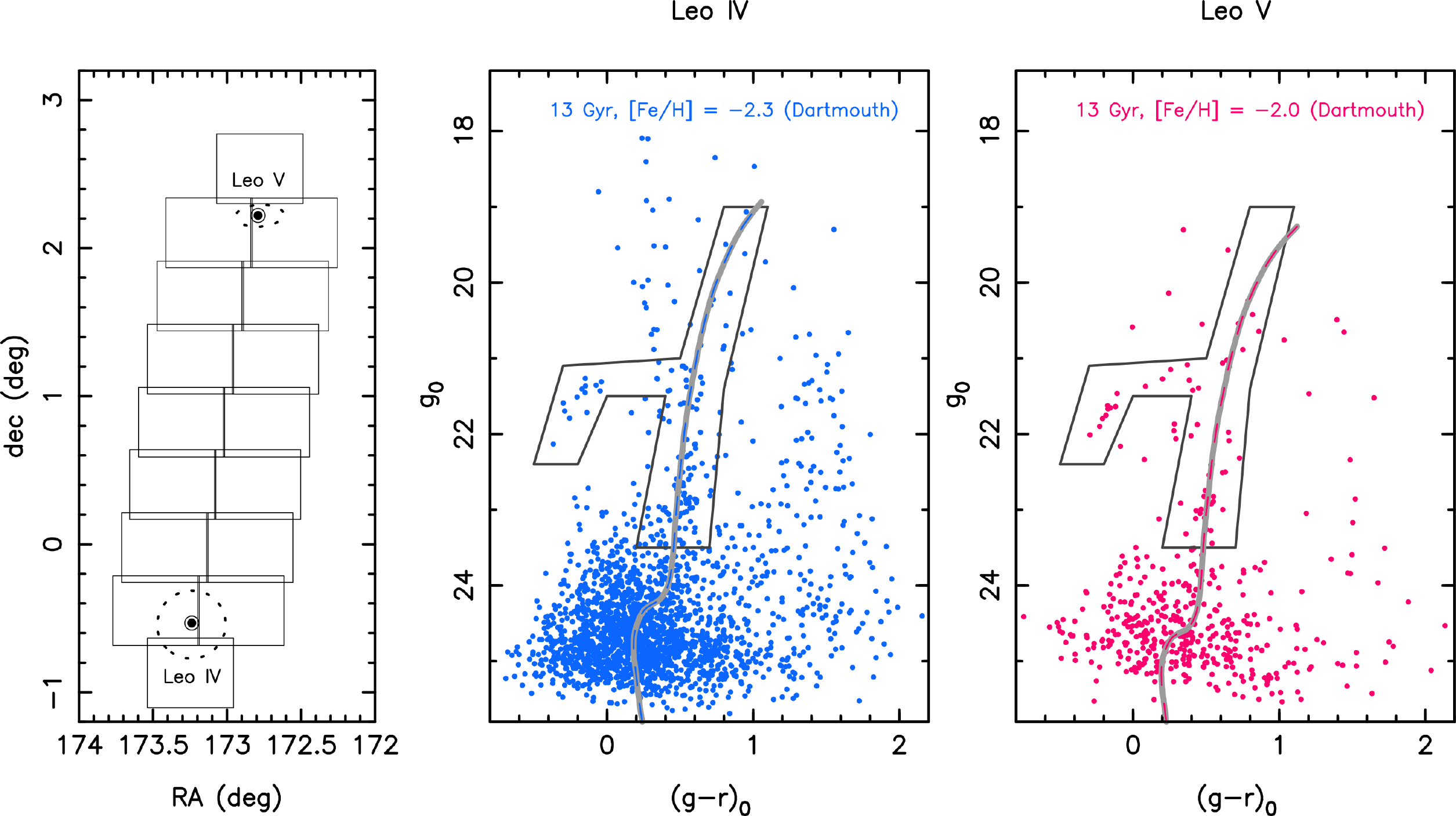}}
\caption{Left panel: footprint of the observed Subaru fields. Locations of the individual Subaru/Suprime-Cam fields are shown, together with dotted ellipses marking 4 half-light radii of Leo IV and Leo V. Middle panel: colour-magnitude diagram of stars within 2 half-light radii of Leo IV, overlaid with a 13-Gyr Dartmouth isochrone with $\mathrm{[Fe/H]} = -2.3~\mathrm{dex}$. The delineated section (same in both CMDs) is used as a selection box for avoiding Leo IV and Leo V members when exploring the question of a stream's presence within the Subaru fields. Right panel: as for middle panel but for Leo V; the isochrone has been altered to reflect the metallicity differences in the Leo IV and Leo V stellar populations.}	
\label{fig:footprint_Leo_CMDs}
\end{center}
\end{figure}	

\begin{table}
\begin{center}
\begin{tabular}{lcc}
\hline
\multicolumn{1}{c}{Parameter}
& \multicolumn{1}{c}{Leo IV}
& \multicolumn{1}{c}{Leo V}\\
\hline
R.A.\, (J2000)* & $11^{\mathrm{h}}32^{\mathrm{m}}58.1^{\mathrm{s}}$ & $11^{\mathrm{h}} 31^{\mathrm{m}} 08.3^{\mathrm{s}}$\\ 
Dec.\, (J2000)* & $-00\null^\circ 32\symbol{39}11.2\symbol{39}\hskip-1pt\symbol{39} $ & $02\null^\circ 13\symbol{39} 20.9\symbol{39}\hskip-1pt\symbol{39}$\\ 
$\theta$ (deg)* & $5^{+45}_{-45}$ & $30^{+15}_{-22}$ \\
$\epsilon$*  &  $0.03^{+0.18}_{-0.03}$ & $0.15^{+0.34}_{-0.15} $\\
$r_\mathrm{h}$ (arcmin)* & $ 2.3^{+0.3}_{-0.4}$ & $1.8\pm 0.5$\\
$D$ (kpc) & $154\pm5$& $175\pm9$\\
\hline
\end{tabular}
\caption{Structural parameters of Leo IV and Leo V (*updated using the current dataset); from top to bottom: centroid position, position angle, ellipticity, half-light radius and heliocentric distance \citep{2009ApJ...699L.125M,2010ApJ...710.1664D}.}
\label{tab:struct_params}
\end{center}
\end{table}

Altogether, 16 Subaru/Suprime-Cam fields were used to obtain a wide coverage of Leo IV, Leo V, and their surrounding environment. Figure \ref{fig:footprint_Leo_CMDs} provides an overview of the location of the observed fields, as well as the colour-magnitude diagrams (CMDs) of the two Leo systems; the data have been photometrically calibrated field-by-field to the SDSS, and extinction-corrected following the \cite{1998ApJ...500..525S} dust maps. The updated structural parameters, derived using a maximum-likelihood procedure as described by \cite{2008ApJ...684.1075M}, are summarised in Table \ref{tab:struct_params}.

\section{Positioning the stellar stream: near or far?} 
\label{sec:analysis}

\begin{figure}
\begin{center}
\vspace{-1.2cm}
\resizebox{0.9\textwidth}{!}{\includegraphics{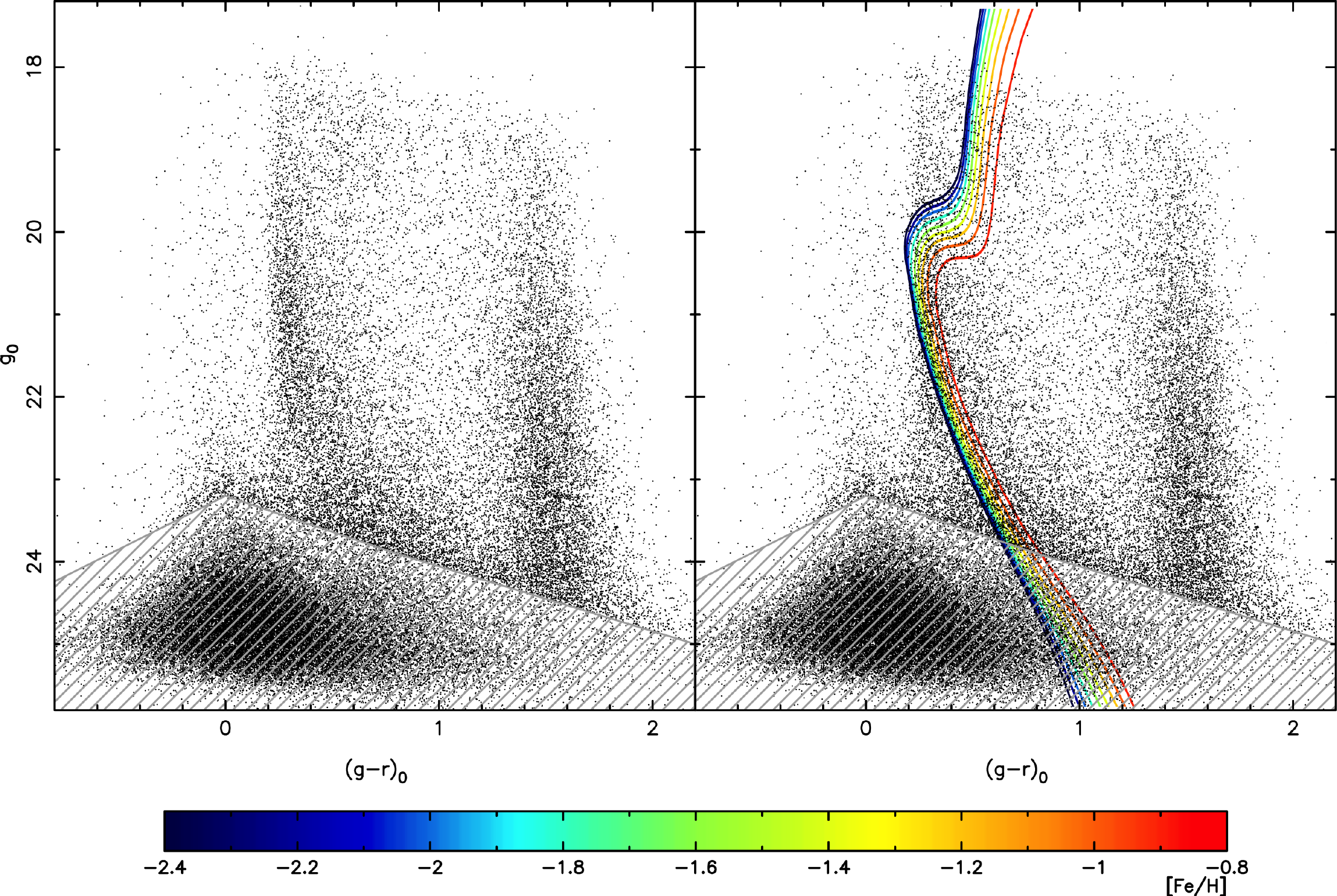}}
\caption{Left panel: stacked CMD of all stars within the Subaru fields. Stellar contributions from Leo IV and Leo V have been removed by designating an elliptical exclusion region with a semi-major axis of 4 half-light radii for each galaxy, together with the selection box shown in Figure~\ref{fig:footprint_Leo_CMDs}. The grey shaded region is expected to contain mainly galaxy contamination; only one in four data points have therefore been plotted. Right panel: as in left panel, but with 13-Gyr Dartmouth isochrones of different metallicities \citep{2008ApJS..178...89D} overlaid to roughly match the location of the main-sequence turn-off feature.}
\label{fig:CMD_stream_isochrones}
\vspace{-0.2cm}
\end{center}
\end{figure}

Despite the improved depth and coverage of the Subaru data with respect to our earlier dataset, we identified no clear spatial overdensity indicative of a distant stellar stream in our new dataset.  However, a CMD of the data from all 16 Suprime-Cam fields, with stars belonging to the dwarf galaxies removed, shows a feature resembling a main-sequence turn-off at $g_0\sim20~\mathrm{mag}$ (see Figure~\ref{fig:CMD_stream_isochrones}). Its location and profile are well matched by a 13-Gyr-old stellar population roughly $20~\mathrm{kpc}$ away with [Fe/H] of approximately $-1.6~\mathrm{dex}$ when compared with Dartmouth isochrones \citep{2008ApJS..178...89D}.

As can be seen from the Milky Way's `field of streams'  \citep{2007ApJ...658..337B}, the stellar overdensity lying on top of Leo IV and V is found near a region of the sky occupied by several prominent streams and stellar overdensities.  Given its currently understood properties and angular separation from the Sagittarius stream, an association of the newly detected structure with Sagittarius is unlikely.  A connection with the nearby Virgo overdensity is in fact more probable, given a close agreement in distance and metallicity \citep[e.g.,][]{2006ApJ...636L..97D,2008ApJ...673..864J}. Whatever its allegiance, the stars once thought to possibly be associated with the Leos are clearly not occupants of the distant halo, but more likely members of a structure enjoying inner-halo residency.

\section{Summary}
\label{sec:summary}

Using deep Subaru/Suprime-Cam data, we have obtained an improved set of structural parameters for the Milky Way dwarf galaxies Leo IV and Leo V. Additionally and more interestingly, we find that a tentative stream of stars previously detected between the two dwarfs is part of a more extended stellar overdensity, coincidentally overlapping the satellite galaxies.  Previously taken as a hint of a feature related to the dwarf galaxies, the deeper and wider Suprime-Cam data reveal a main-sequence turn-off in the stacked CMD consistent with that of a 13-Gyr-old stellar population with [Fe/H] of $-1.6$~dex at a distance of approximately 20~kpc. 

While being far enough away from the multiple wraps of the imposing Sagittarius stellar stream to avoid this affiliation, the on-sky proximity of the stellar overdensity enveloping Leo IV and Leo V to the Virgo stellar overdensity, along with a comparable distance and consistent metallicity, appears to imply an association with the latter structure. Indeed, our former Leo stellar-bridge hopeful appears to be suffering a constellation membership crisis. Instead of being trapped inside the lion's den, however, the stars of the newly detected stellar substructure manage to escape the Leos' jaws, only to be lured into the arms of Virgo.

\acknowledgements
SJ thanks the Alexander von Humboldt Foundation for a research fellowship held during the time this work was carried out. BC is an Alexander von Humboldt research fellow.


\begin{thebibliography}{}
\expandafter\ifx\csname natexlab\endcsname\relax\def\natexlab#1{#1}\fi
\expandafter\ifx\csname url\endcsname\relax
  \def\url#1{\texttt{#1}}\fi
\expandafter\ifx\csname urlprefix\endcsname\relax\def\urlprefix{URL }\fi
\providecommand{\eprint}[2][]{\url{#2}}

\bibitem[{{Belokurov} et~al.(2007{\natexlab{a}})}]{2007ApJ...658..337B}
{Belokurov}, V., {Evans}, N.~W., {Irwin}, M.~J. {et al.}
  2007{\natexlab{a}}, \apj, 658, 337

\bibitem[{{Belokurov} et~al.(2007{\natexlab{b}})}]{2007ApJ...654..897B}
{Belokurov}, V., {Zucker}, D.~B., {Evans}, N.~W. {et al.}
  2007{\natexlab{b}}, \apj, 654, 897

\bibitem[{{Belokurov} et~al.(2008)}]{2008ApJ...686L..83B}
{Belokurov}, V., {Walker}, M.~G., {Evans}, N.~W. {et al.}
2008, \apjl, 686, L83

\bibitem[{{de Jong} et~al.(2010)}]{2010ApJ...710.1664D}
{de Jong}, J.~T.~A., {Martin}, N.~F., {Rix}, H.-W. {et al.}
2010, \apj, 710, 1664

\bibitem[{{Dotter} et~al.(2008)}]{2008ApJS..178...89D}
{Dotter}, A., {Chaboyer}, B., {Jevremovi{\'c}}, D. {et al.}
2008, \apjs, 178, 89

\bibitem[{{Duffau} et~al.(2006)}]{2006ApJ...636L..97D}
{Duffau}, S., {Zinn}, R., {Vivas}, A.~K.  {et al.}
2006, \apjl, 636, L97

\bibitem[{{Juri{\'c}} et~al.(2008)}]{2008ApJ...673..864J}
{Juri{\'c}}, M., {Ivezi{\'c}}, Z., {Brooks}, A. {et al.}
2008, \apj, 673, 864

\bibitem[{{Martin} et~al.(2008)}]{2008ApJ...684.1075M}
{Martin}, N.~F., {de Jong}, J.~T.~A., \& {Rix}, H.-W. 
2008, \apj, 684, 1075

\bibitem[{{Moretti} et~al.(2009)}]{2009ApJ...699L.125M}
{Moretti}, M.~I., {Dall'Ora}, M., {Ripepi}, V. {et al.}
  2009, \apjl, 699, L125

\bibitem[{{Schlegel} et~al.(1998)}]{1998ApJ...500..525S}
{Schlegel}, D.~J., {Finkbeiner}, D.~P. \& {Davis}, M. 1998, \apj, 500, 525

\bibitem[{{York} et~al.(2000)}]{2000AJ....120.1579Y}
{York}, D.~G. \& {The SDSS Collaboration} 2000, \aj, 120, 1579

\end{thebibliography}
\end{document}